# Experimental study of the structure of a lean premixed indane/CH$_4$/O$_2$/Ar flame

E. Pousse, P.A. Glaude, R. Fournet, F.Battin-Leclerc

*Département de Chimie-Physique des Réactions, UMR 7630 CNRS Nancy Université,*

*1 rue Grandville, BP 451, 54001 Nancy, France*

*pierre-alexandre.glaude@ensic.inpl-nancy.fr*

**ABSTRACT**

In order to better understand the chemistry involved during the combustion of components of diesel fuel, the structure of a laminar lean premixed methane flame doped with indane has been investigated. The gases of this flame contains 7.1% (molar) of methane, 36.8% of oxygen and 0.90% of indane corresponding to an equivalence ratio of 0.74 and a ratio C$_9$H$_{10}$/CH$_4$ of 12.75%. The flame has been stabilized on a burner at a pressure of 6.7 kPa using argon as diluant, with a gas velocity at the burner of 49.2 cm/s at 333 K. Quantified species included usual methane C$_0$-C$_2$ combustion products, but also 11 C$_3$-C$_5$ hydrocarbons and 3 C$_1$-C$_3$ oxygenated compounds, as well as 17 aromatic products, namely benzene, toluene, phenylacetylene, styrene, ethylbenzene, xylenes, trimethylbenzenes, ethyltoluenes, indene methylindane, methylindene, naphthalene, phenol, benzaldehyde, benzofuran. The temperature was measured thanks to a thermocouple in PtRh (6%)-PtRh (30%) settled inside the enclosure and ranged from 800 K close to the burner up to 2000 K in the burned gases.

**INTRODUCTION**

If many detailed kinetic models are available for the oxidation of mixtures representative of gasolines, they are much less numerous in the case of diesel fuels because of their more complex composition. The constituents of diesel fuel contain from 10 to 20 carbon atoms and include about 30% (mass) of alkanes, the remaining part being mainly alkylcyclohexanes (24%), alkyldecalines (15%), alkylbenzenes (10%), indanes (7%) and polycyclic naphtenoaromatic compounds [1]. If the oxidation of alkanes has been extensively studied, the abundance of models diminishes considerably when other families of components are considered and very few models exist for substituted cycloalkanes and aromatics compounds, except from toluene [2]. The oxidation of alkylbenzenes with alkylic side-chains from C$_2$ to C$_4$ has been studied in a flow reactor in Princeton [3] [4] [5] and that of n-propylbenzene in a jet-stirred reactor at Orleans [6]. The autoignition of alkylbenzenes (toluene, xylenes, ethylbenzene, trimethylbenzenes, propylbenzene, ethyltoluenes and butylbenzene) has also been investigated in the rapid compression machine of Lille [7] [8]. The oxidation of indane has only been studied in a jet stirred reactor [9] at Orleans. No measurement in flame has yet been reported for this species.

The purpose of the present paper is to experimentally investigate the structure of a premixed laminar methane flame containing indane. The use of a methane flame will allow us to have a reactive mixture rich in methyl radicals and to be more representative of combustion mixtures containing larger hydrocarbons than hydrogen or C$_2$ flames. This study will be performed using a lean flame for the chemistry to be better representative of that occurring in engines controlled via Homogenous Charge Combustion Ignition (HCCI) which are under development. These results will be use in order to develop a new mechanism for the oxidation of indane based on our experience in modelling the reactions of both alkanes [10] and light aromatic compounds (benzene [11] and toluene [12]).

**EXPERIMENTAL PROCEDURE**

The experiments were performed using an apparatus developed in our laboratory to study temperature and stable species profiles in a laminar premixed flat flame at low pressure and recently used in the case of rich methane flames doped by light unsaturated soot precursors [13] [14] [15].
The body of the flat flame matrix burner, provided by McKenna Products, was made of stainless steel, with an outer diameter of 120 mm and a height of 60 mm (without gas/water connectors). This burner was built with a bronze disk (95% copper, 5% tin). The porous plate (60 mm diameter) for flame stabilization was water cooled (water temperature: 333 K) with a cooling coil sintered into the plate. The burner could be operated with an annular co-flow of argon to favor the stabilization of the flame.
This horizontal burner was housed in a water-cooled vacuum chamber evacuated by two primary pumps and maintained at 6.7 kPa by a regulation valve. This chamber was equipped of four quartz windows for an optical access, a pressure transducer (MKS 0-100 Torr), a microprobe for samples taking and a thermocouple for temperature measurements. The burner could be vertically translated, while the housing and its equipments were kept fixed. A sighting telescope measured the position of the burner relative to the probe or the thermocouple with an accuracy of 0.01 mm. The flame was lighted on using an electrical discharge.
Gas flow rates were regulated by RDM 280 Alphagaz and Bronkhorst (El-Flow) mass flow regulators. Methane (99.95 % pure) was supplied by Alphagaz - Air Liquide. Oxygen (99.5% pure) and argon (99.995% pure) were supplied by Messer. Liquid indane, supplied by Alfa Aesar (purity 99%), was contained in a glass vessel pressurized with argon. After each load of the vessel, argon bubbling and vacuum pumping were performed in order to remove oxygen traces dissolved in the liquid hydrocarbon fuel. The liquid reactant flow rate was controlled by using a liquid mass flow controller, mixed to the carrier gas and then evaporated by passing through a single pass heat exchanger, the temperature of which was set above the boiling point of the mixture. Carrier gas flow rate was controlled by a gas mass flow controller located before the mixing chamber.
Temperature profiles were obtained using a PtRh (6%)-PtRh (30%) type B thermocouple (diameter 200 µm). The thermocouple wire was sustained by a fork and crosses the flame horizontally to avoid conduction heat losses. The junction was located at the centre of the burner. The thermocouple was coated with an inert layer of $BeO-Y_2O_3$ to prevent catalytic effects [16]. The ceramic layer was obtained by damping in a hot solution of $Y_2(CO_3)_3$ (93% mass.) and BeO (7% mass.) followed by a drying in a Meker burner flame. This process was reiterated (about ten times) until the whole metal was covered. Radiative heat losses are corrected using the electric compensation method [17].
The sampling probe was in quartz with a hole of about 50 µm diameter ($d_i$). The probe was finished by a small cone with an angle to the vertical of about 20°. For temperature measurements in the flames perturbed by the probe, the distance between the junction of the thermocouple and the end of the probe was taken equal to two times $d_i$, i.e. to about 100 µm. The sampling quartz probe was directly connected to a heated passivated transfer line heated at 423 K. This line was itself connected to a heated pressure transducer (MKS 0-100 Torr) and through heated valves to a turbomolecular pomp, a pyrex line and a heated stainless steel loop located inside a gas chromatograph. The pressure in the transfer line was always below 1.3 kPa so that the pressure drop between the flame and the inlet of the probe ensured reactions to be frozen. The on-line connection to a chromatograph via a heated transfer line was found necessary in order to analyse compounds above $C_6$.

The analyses could be then performed according to two methods:
- ➢ Gas samples of compounds with a sufficient vapour pressure were directed through the pyrex line towards a volume which was previously evacuated by the turbo molecular pump down to $10^{-7}$ kPa and which was then filled up to a pressure of 1.3 kPa (pressures in the pyrex line were measured by a MKS 0-10 Torr pressure transducer before compression and by a MKS 0-1000 Torr one after compression) and collected in a pyrex loop after compression by a factor 5 by using a column with rising mercury pressure. A chromatograph with a Carbosphere packed column and helium or argon as carrier gas was used to analyse $O_2$, $H_2$, $CO$ and $CO_2$ by thermal conductivity detection and $C_2H_2$, $C_2H_4$, $C_2H_6$ by flame ionisation detection (FID). Water was detected by TCD but not quantitatively analyzed. Calibrations were performed by analysing a range of samples containing known pressures of each pure compound to quantify and mole fractions were derived from the known total pressure in the pyrex loop. This method of sampling could allow us to analyse heavier hydrocarbons from allene to toluene using a chromatograph with a Haysep packed column with FID and nitrogen as gas carrier gas as in our previous study [15], but it was found more accurate and faster to use the second method. When both methods were used for these compounds, a very good agreement was obtained between both of them.
- ➢ Gas samples were directed towards the loop located inside the gas chromatograph and previously evacuated by the turbo molecular pump down to $10^{-7}$ kPa. The loop was then filled up to a pressure of 1.3 kPa and its content was just after analyzed using FID and helium as gas carrier gas by using a capillary column (a HP-Plot Q or a HP-1 column). This method of sampling was used to analyze methane, all the hydrocarbons from $C_3$ and the oxygenated products other than carbon oxides.

**RESULTS AND DISCUSSION**

A laminar premixed flat flame has been stabilized on the burner at 6.7 kPa (50 Torr) with a gas flow rate of 5.44 l/min corresponding to a gas velocity at the burner of 49.2 cm/s at 340 K and with mixtures containing 7.1% (molar) of methane, 36.8% of oxygen and 0.90% of indane corresponding to an equivalence ratio of 0.74.

Figure 1 displays the experimental temperature profiles obtained with and without the probe showing that the presence of the probe induces a thermal perturbation involving a lower temperature. Without the probe, the lowest temperatures measured the closest to the burner (0.4 mm above) were around 1115 K. Due to the thinness of this lean flame and the size of the thermocouple, it was not possible to measure lower temperatures. The highest temperatures were reached from 6 mm above the burner and were around 1995 K.

FIGURE 1

Figure 2 presents the profiles of both hydrocarbons reactants and shows that indane is completely consumed close to the burner, at 2 mm height, while some methane remains up to 3 mm.

FIGURE 2

Figures 3 present the profiles of oxygen (fig. 3a), hydrogen (fig. 3a), water (fig. 3a) and of carbon monoxide (fig. 3b) and carbon dioxide (fig. 3b) above the burner. The mole fraction of water has been obtained from a material balance on the other major species. In this lean flame, the profiles of carbon monoxide and hydrogen display a marked maximum at 2.5 mm height and the major final products are for a large extent only carbon dioxide and water.

FIGURE 3

Figure 4 present the profiles of $C_2$ species. Acetylene is the most abundant $C_2$ species (peak mole fraction of 1800 ppm) and is produced last. It reaches its maximum concentration close to the burner around 2.5 mm. Ethylene is slightly less abundant than acetylene. The profiles of ethylene and ethane peak around 2 mm.

FIGURE 4

Figure 5 presents the profiles of the observed $C_3$ products, with propene (fig. 5a) and propane (fig. 5a) peaking first around 2 mm above the burner, while the maxima for allene (fig. 5b) and propyne (fig 5b) are around 2.2 mm. The peak mole fraction observed for propene (80 ppm) is close to those of propyne and allene but it's more than ten times larger than that of propane.

FIGURE 5

Figure 6 displays the profiles of $C_4$ species. Butenes (fig. 6a) are produced first and reach its maximum concentration close to the burner, around 1.9 mm. The profiles of diacetylene (fig. 6b), vinylacetylene (fig. 6b) and butadienes (fig. 6b) peak around 2.2 mm. The most abundant $C_4$ compounds are vinylacetylene and 1,3-butadiene with peak mole fractions of 70 and 100 ppm respectively.

FIGURE 6

Figure 7 presents the profiles of cyclopentadiene (CPTD) and methylcyclopentadiene (MCPTD). The profiles of methylcyclopentadiene peak further around 1.8 mm and the maxima of that of cyclopentadiene, the most abundant specie with a peak mole fraction of 70 ppm, is around 2.2 mm.

FIGURE 7

Figure 8 displays the profiles of light oxygenated species: methanol, acetaldehyde and acroleïn which are intermediate products of the combustion of methane. Methanol and acetaldehyde, which are the most abundant species with peak mole fractions of 160 and 60 ppm respectively, are produced early and reach their maximum concentration close to the burner, around 1.9 mm. The profiles of acroleïn peak around 2.1 mm.

FIGURE 8

Figure 9 present the profiles of monocyclic aromatic products. The maxima of the profiles of, ethylbenzene (fig. 9c) and ethyltoluene (fig. 9d) occur first around 1.8 mm above the burner, followed by that of toluene (fig. 9a), xylenes (fig. 9b) and trimethylbenzene (fig. 9d) around 2.0 mm and finally that of benzene (fig. 9a), styrene (fig. 9c) and phenylacetylene (fig. 9c) around 2.3 mm. The most abundant of these species are benzene and toluene, with peak mole fractions of 900 and 600 ppm respectively, and to a lesser extent styrene (with 250 ppm), ethylbenzene (with 200 ppm) and phenylacetylene (with 100 ppm).

FIGURE 9

Figure 10 displays the profiles of bicyclic aromatic compounds. The maxima of the profiles of indene (fig. 10a) and methylindane (fig. 10c) are observed first around 1.7 mm above the burner, followed by that of methylindene (fig. 10c) and benzocyclobutane (fig. 10b) around 2.2 mm and finally that of naphthalene (fig. 10c) around 2.3 mm. The most abundant of these species is indene. Indene is the major intermediate $C_2+$ species detected in this flame with a peak mole fraction of 1600 ppm.

FIGURE 10

Finally figure 11 presents the profiles of oxygenated aromatic species. Benzaldehyde, the most abundant of these species (peak mole fraction of 80 ppm), reach its maxima first around 2.0 mm above the burner. The profiles of phenol peak further around 2.2 mm and the maximum of that of benzofuran is the last one around 2.3 mm.

FIGURE 11

With the addition of indane, the formation of large amounts of acetylene, ethylene, indene, benzene, toluene, styrene, ethylbenzene and phenylacetylene were observed. The most important ways of decomposition of indane is probably H-atom abstraction by metatheses or by bimolecular initiation with oxygen yields 1-indanyl and 2-indanyl. The formation of 1-indanyl should be favored on the basis of the number of available abstraction sites (2 $CH_2$ groups) and the fact that this radical is stabilized by resonance. β-scission C-C bond in 1-indanyl followed by addition of hydrogen or methyl rdical $CH_3$ can yields respectively α-methylstyrene or α-ethylstyrene (wich experimentally observed but not quantified). Thermal decomposition of indanyl radicals can also led to acetylene and benzyl radical. Benzyl yields toluene via hydrogen addition, whereas its oxidation yields benzaldehyde. The recombination of benzyl and methyl radicals yields ethylbenzene. Indanyl radicals also can decomposate to benzene and propargyl radical ($C_3H_3$) or to phenylacetylene and $CH_3$. Indene is formed via β-scission C-H bond in 1-indanyl and 2-indanyl. Its probably the most important voice decomposition of indanyl radicals by the fact that indene is the most intermediaire important specie experimentally observed. Combinations of 1-indanyl with methyl radical $CH_3$ yields 1-methylindane. Indene oxidation yields 1-indanone whereas H atom abstraction on indene yields indenyl. Indenyl oxidation yields indenonyl that can rearrange to eliminate CO and 2-phenylvinyl that in turn decomposes to form benzene, or forms styrene by hydrogenation. Indenyl can rearrange to yield α-ethynylbenzyl that,by oxidation yields benzaldehyde.

**CONCLUSION**

This paper presents new experimental results for a lean premixed laminar flame of methane seeded with indane which can be considered as an important component of diesel fuels. Profiles of temperature have been measured and mole fraction profiles have been obtained for 41 identified stable species from $C_0$ to $C_{10}$, including 17 aromatic products and 6 oxygenated compounds other than the reactants.

**FIGURE CAPTIONS**

Figure 1: Temperature profiles: experimental measurements performed without and with the sampling probe.
Figure 2: Profiles of the mole fractions of both hydrocarbon reactants.
Figure 3: Profiles of the mole fractions of oxygen and the main oxygenated products.
Figure 4: Profiles of the mole fractions of $C_2$ species.
Figure 5: Profiles of the mole fractions of $C_3$ species.
Figure 6: Profiles of the mole fractions of $C_4$ species.
Figure 7: Profiles of the mole fractions of cyclopentadiene and methylcyclopentadiene.
Figure 8: Profiles of the mole fractions of oxygenated $C_1$-$C_3$ species.
Figure 9: Profiles of the mole fractions of $C_6$-$C_{10}$ monoaromatic species.
Figure 10: Profiles of the mole fractions of biaromatic species.
Figure 11: Profiles of the mole fractions of oxygenated aromatic species.
Figure 12: Proposed major pathways in the oxidation of indane in flame

Figure 1

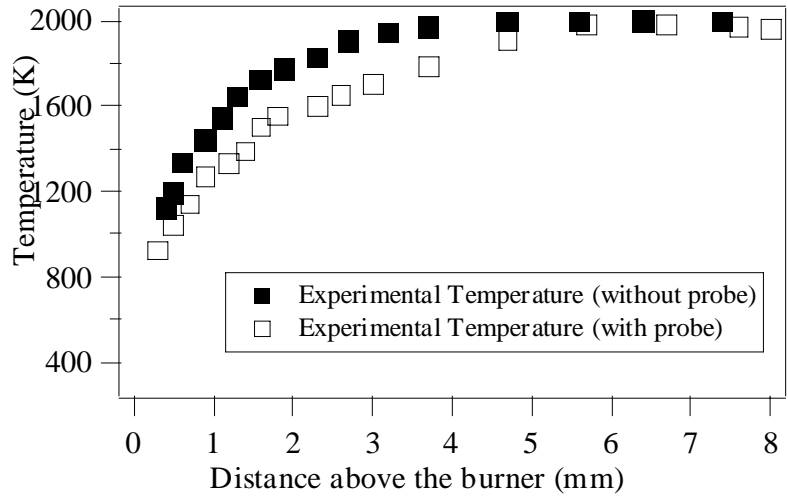

Figure 2

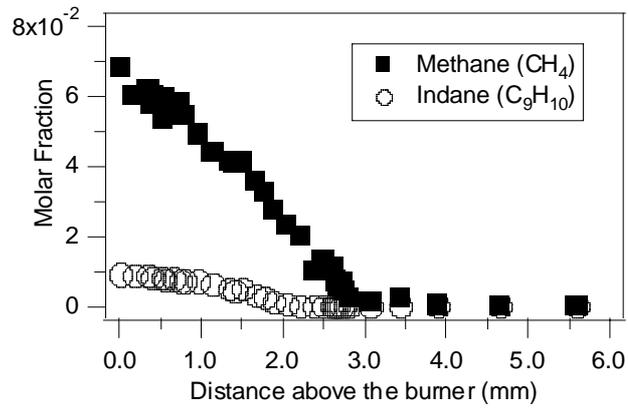

Figure 3

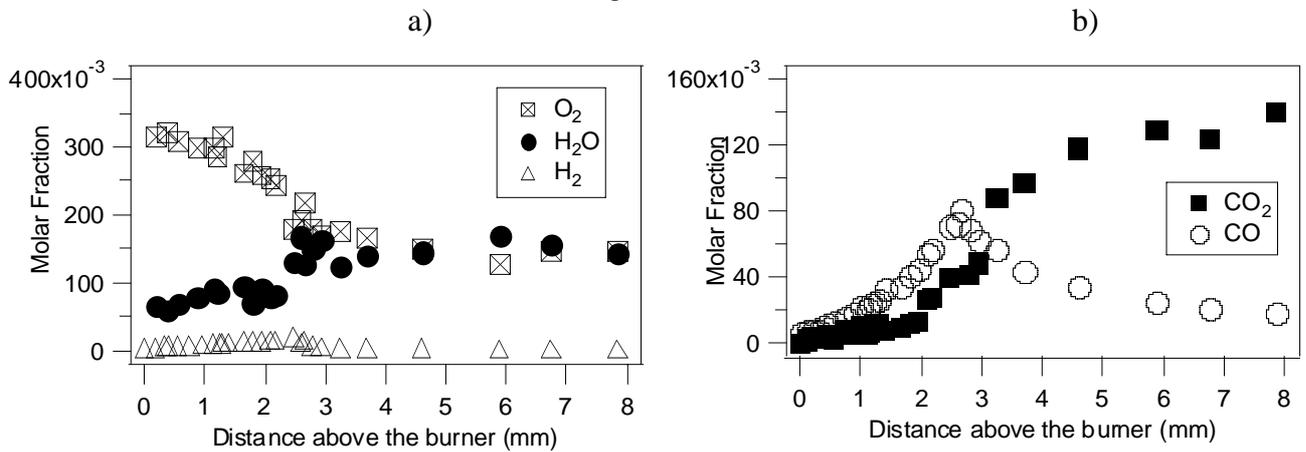

Figure 4

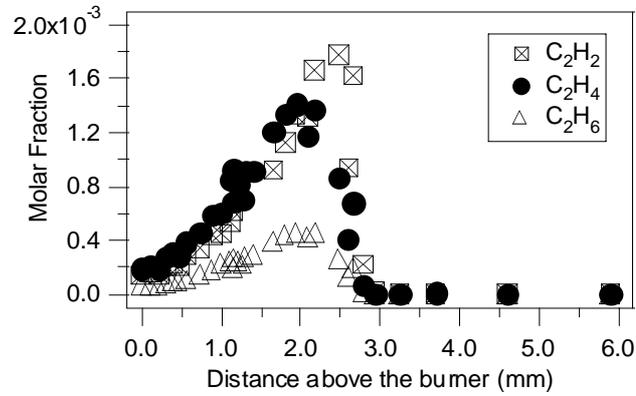

Figure 5

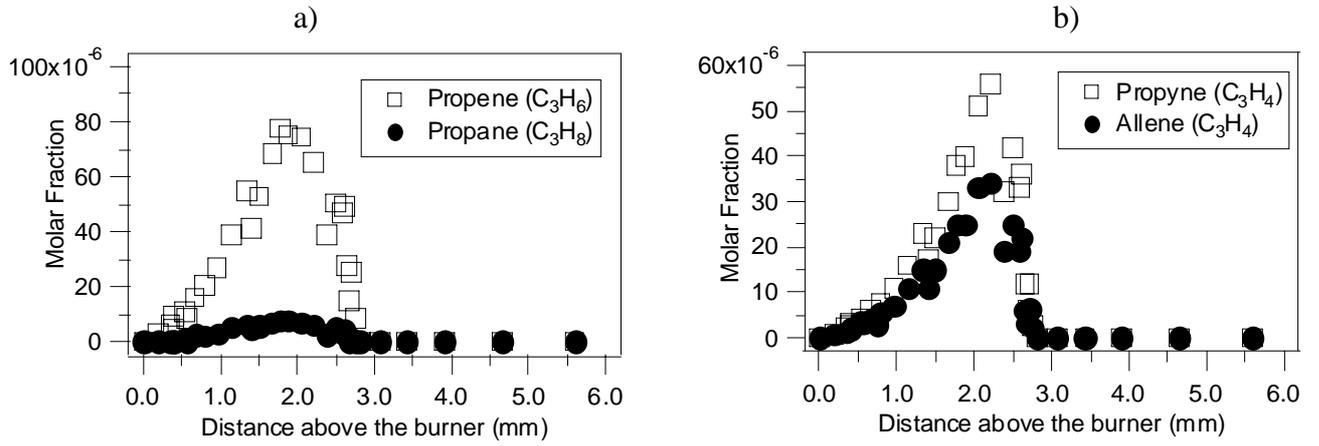

Figure 6

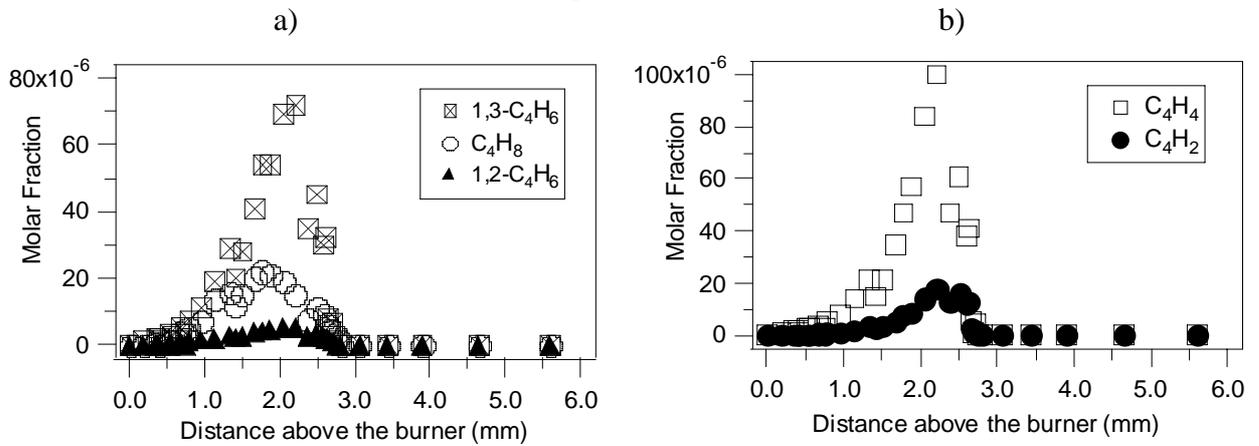

Figure 7

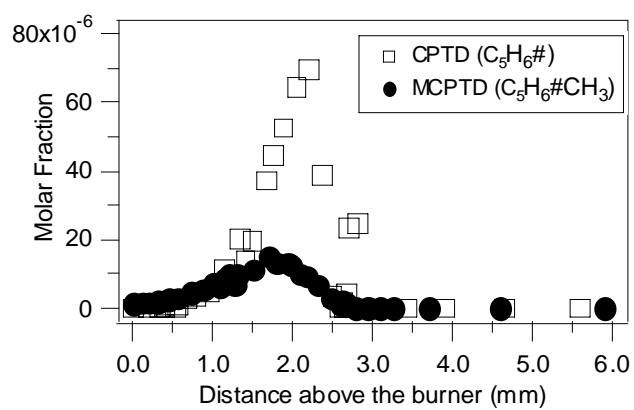

Figure 8

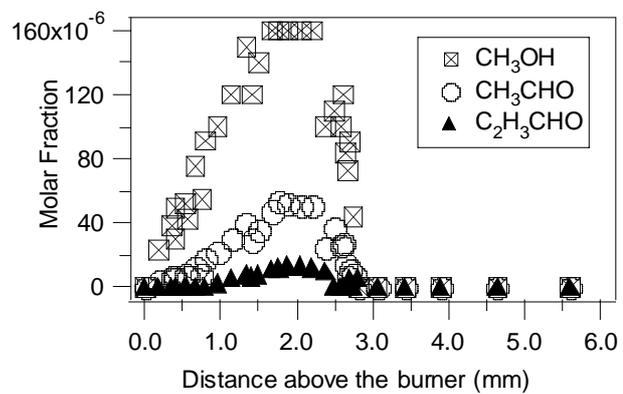

Figure 9

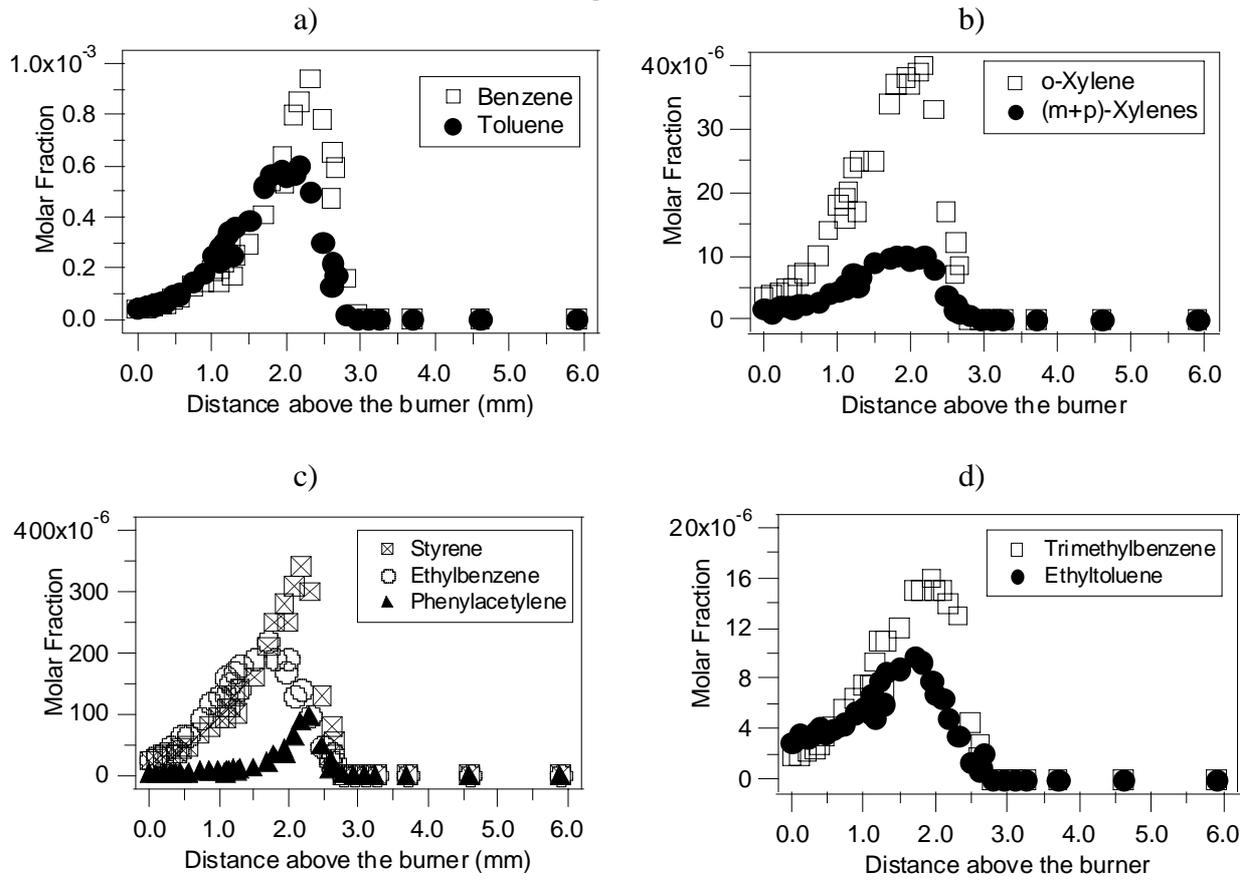

Figure 10

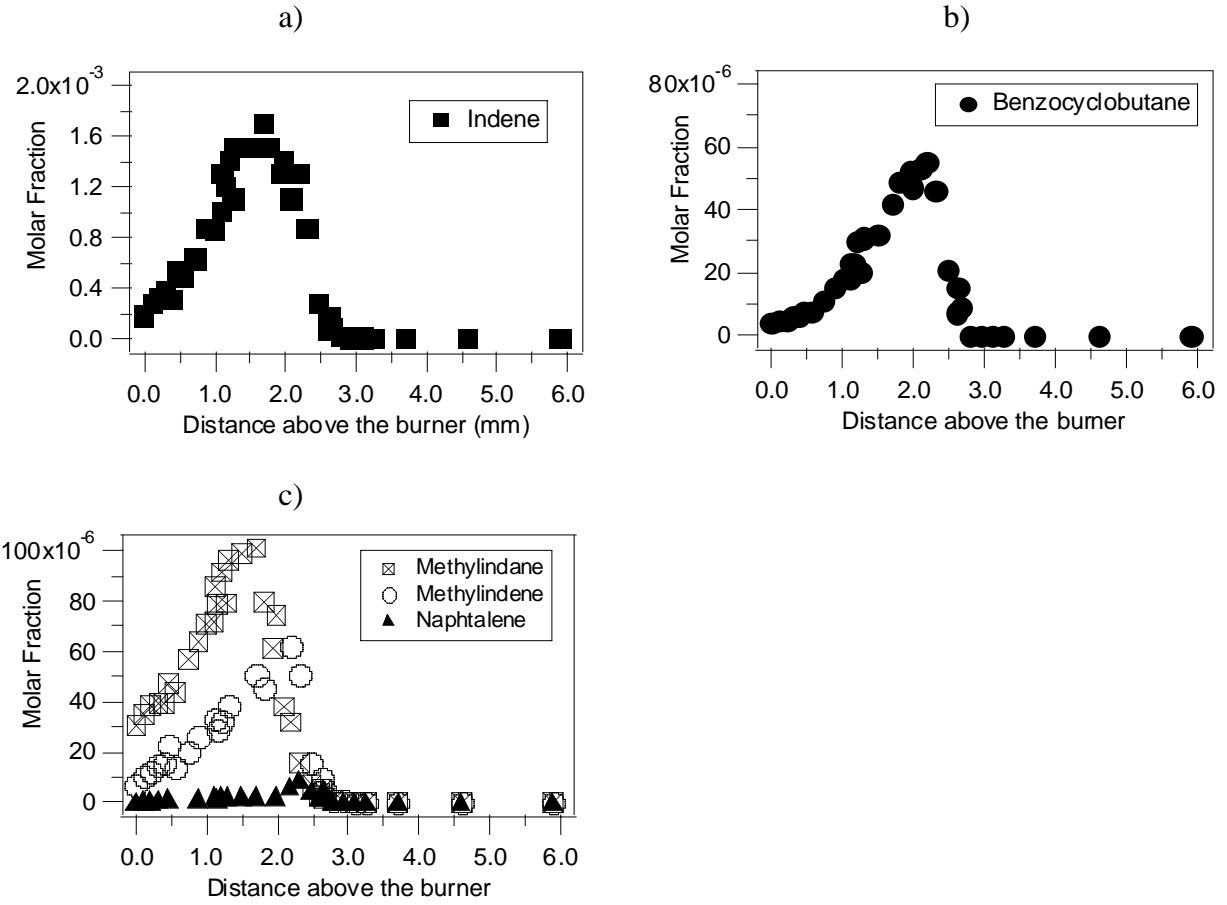

Figure 11

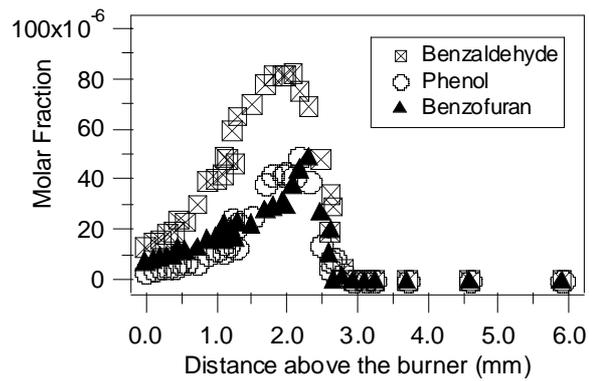

Figure 12

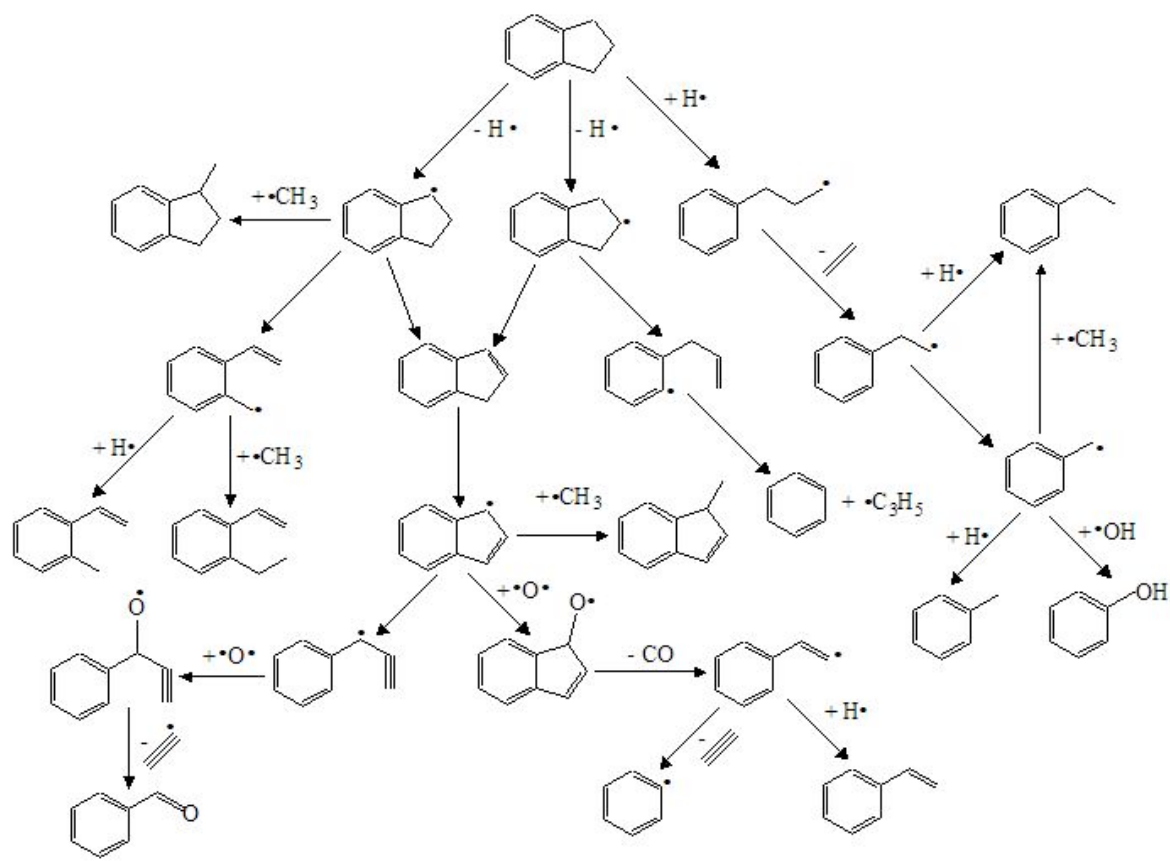